# A modified W-W interatomic potential based on *ab initio* calculations

Jun Wang, Yulu Zhou, Min Li, and Qing Hou[*]

*Key Laboratory for Radiation Physics and Technology, Institute of Nuclear Science and Technology, Sichuan University, Chengdu 610064, People's Republic of China*

[*]Corresponding author. E-mail addresses: qhouscu@163.com (Q. Hou)

Tel: 0086-028-85412104, Fax: 0086-028-85410252

## ABSTRACT

Based on *ab initio* calculations, a Finnis-Sinclair-type interatomic potential for W-W interactions has been developed. The modified potential is able to reproduce the correct formation energies of self-interstitial atom defects in tungsten, offering a significant improvement over the Ackland-Thetford tungsten potential. Using the modified potential, the thermal expansion is calculated in a temperature range from 0 K to 3500 K, and the results are in reasonable agreement with the experimental data, thus overcoming the shortcoming of the negative thermal expansion using the Derlet-Nguyen-Manh-Dudarev tungsten potential. The present W-W potential is also applied to study in detail the diffusion of self-interstitial atoms in tungsten. It is revealed that the initial self-interstitial atom initiates a sequence of W atom displacements and replacements in the $\langle 111 \rangle$ direction. An Arrhenius fit to the diffusion data at temperatures below 550 K indicates a migration energy of 0.022 eV, in reasonable agreement with the experimental data.

# 1. Introduction

Due to its high melting point, high thermal conductivity and low sputtering erosion, tungsten is a promising candidate for the plasma facing materials in fusion applications, e.g., for the divertor armor of the future thermonuclear fusion reactors [1,2]. It will be subjected to high-energy neutron irradiation, which produces vacancies and self-interstitial atoms (SIAs) due to atomic displacements and generates helium by transmutation reactions, throughout its lifetime. The accumulation and interaction of these defects in materials is believed to produce swelling, cause embrittlement, and induce blisters, ultimately resulting in the deterioration of mechanical properties of the material. Understanding and predicting these important physical processes has been a subject of considerable interest [3-12].

Molecular dynamics (MD) simulations, as one of the effective and powerful tools in materials science, are extensively used to study many physical processes, such as defect formation and its influence on the mechanical properties of materials. Of critical importance for the reliable performance of MD simulations is an appropriate description of the interatomic potentials. Hence, significant effort has been dedicated to developing accurate W-W potentials, such as the Finnis-Sinclair (FS) potential [13], the embedded atom method (EAM) [14], the modified embedded atom method (MEAM) [15], the bond-order potential with Tersoff-Brenner-type form(BOP) [16], and the bond-order potential based on the tight-binding approximation (TB-BOP) [17]. Among these potentials, the FS or FS-type potential is very attractive, due to its simple analytic formalism and high computational efficiency. The most commonly used potential is the Ackland-Thetford potential (AT) [18], which was developed from the FS potential [13] and is able to predict a realistic pressure-volume relationship; however, compared to the *ab initio* calculations, the obtained SIA formation energies are entirely too low and the energy differences between various SIA configurations is quite different from those of the *ab initio* calculations [9,10]. Recently, Derlet et al. developed a Finnis-Sinclair-type potential (here denoted DND) that is fitted to the *ab initio* calculations and is able to reproduce the correct energies of various SIA configurations in tungsten [19]; however, it predicts negative thermal expansion when the temperature is above approximately 300 K, in conflict with the experimental results [5,20,21].

Molecular dynamics simulations performed by Fikar et al. demonstrated that different W-W potentials result in different defect behavior [4,5]. Our previous study also indicated that different W-W potentials have a significant influence on the He defect properties in tungsten [22]. Due to the shortcomings of the AT and DND potentials mentioned above, a more reliable W-W potential is necessary to investigate the defect behavior in tungsten in a more accurate way. In this paper, we propose a FS-type interatomic potential for W-W interactions on the basis of *ab initio* calculations to investigate the defect properties and SIA diffusion behavior in W.

## 2. W-W interaction potential

According to the FS formalism [13], the total energy of an N-atom system is given by

$$E_{tot} = \sum_{i}^{N} F(\rho_i) + \frac{1}{2} \sum_{ij, i \neq j}^{N} V(r_{ij}), \qquad (1)$$

where the embedding energy $F(\rho)$ is given by

$$F(\rho) = -A\sqrt{\rho}, \qquad (2)$$

and

$$\rho_i = \sum_{j, j \neq i} f(r_{ij}). \qquad (3)$$

For the present work, the pairwise potential function and the electronic density function will be represented by a series of polynomial functions,

$$V(r) = \sum_{n=1}^{N} \left( A_n + B_n r + C_n r^2 + D_n r^3 + E_n r^4 + F_n r^5 + G_n r^6 + H_n r^7 \right) \Theta(r_n - r) \Theta(r - r_{n+1}),$$

$$r_1 < r \leq r_{N+1} \qquad (4)$$

and

$$f(r) = \sum_{n=1}^{N'} \left( A'_n + B'_n r + C'_n r^2 + D'_n r^3 + E'_n r^4 + F'_n r^5 + G'_n r^6 + H'_n r^7 \right) \Theta(r'_n - r) \Theta(r - r'_{n+1}),$$

$$r'_1 < r \leq r'_{N+1}, \qquad (5)$$

respectively, where $\Theta(x)$ is a Heaviside step function, defined as $\Theta(x) = 1$ for $x \geq 0$ and $\Theta(x) = 0$ for $x < 0$.

To obtain a usable empirical potential, the potential parameters from these functions given above are optimized using a genetic algorithm [23] combined with MD simulations to reproduce as closely as possible a variety of physical properties for tungsten. At the intersection points, the function values and their first derivatives are guaranteed to be continuous. In the present study, in addition to the basic physical properties (lattice constant, cohesive energy, elastic constants and vacancy formation), the SIA defect formation energies from the *ab initio* calculations [10] are also included, as listed in Table 1. Considering computer capacity, all simulations were performed for 432 atoms in pure bulk, adding or removing atoms as needed for the defect configuration. For all simulations, periodic boundary conditions and constant volume were used. Because the process for determining the optimal parameters is a rather complex process, a manual interruption is often required to produce the desired properties. While a set of optimal parameters are obtained, the basic properties (such as SIA formation energies) and other properties (such as thermal expansion) that are not included in the fitting are examined for the purpose of obtaining a more reliable potential. The resulting parameters are presented in Table 2. For reference, Fig. 1 displays the electronic density function multiplied by $A^2$ and pairwise potential radial function for AT, DND and the present tungsten potentials. As seen in Table 1, the basic physical properties and SIA defect properties are reproduced well. The $\langle 111 \rangle$ dumbbell is determined to be the most stable configuration in tungsten, its formation energy is 9.58 eV, and the difference with the $\langle 110 \rangle$ dumbbell is 0.28 eV, in excellent agreement with the *ab initio* calculations [9,10]. The present potential offers a significant improvement over the W-W potential of Ackland et al. [18]. In the following section, we will see that it also improves the DND potential by investigating the thermal expansion.

## 3. Thermal expansion and melting point

Thermal expansion is a critical characteristic property of all materials. Here, the linear thermal expansion is employed and its form is expressed as follows:

$$\beta(T) = \left[ L(T) - L_0 \right] / L_0 \quad , \qquad (6)$$

where $l_0$ and $l$ refer to the specimen length at reference temperature and at the temperature T, respectively. For comparison purposes, we set the reference temperature to be 293.15 K as that in the experiment [20]. By relaxing the tungsten sample containing 2000 W atoms at different temperature using MD simulations, the thermal expansion of three potentials (AT, DND and the present potential) were determined. The temperature of the system was varied between 0 and 3500 K using an electron-phonon coupling model, in which the electron gas is considered as a thermostat at constant temperature. For each temperature, the system volume was adjusted to maintain a pressure of approximately 0 kbar. As shown in Fig. 2, the thermal expansion from the AT potential is significantly higher than that of the experimental results, while for the DND the negative thermal expansion appears, in conflict with the experimental results [20,21]. The thermal expansion is directly related to the interatomic potential of a crystal and is closely related to other thermodynamic and mechanical properties. In particular, high-temperature thermal expansion is most sensitive to the formation of thermal defects [21]. So, reasonably predicting the thermal expansion is an important characteristic of a good many-body atomic potential. The present potential is able to correctly predict the thermal expansion, demonstrating that it is a reliable potential over a fairly broad temperature range.

Further, the melting point was determined using the present potential by performing the MD simulations of a two phase system [16,24,25] with a $20\times10\times10$ simulation box containing 4000 W atoms. Solid and liquid phases were joined and were relaxed to different temperatures. If the temperature is above the melting temperature, the interface moves into the solid region and the sample will melt completely with further evolution; otherwise, crystallization occurs. Here, the melting temperature is $4100\pm50$ K, which is 400 K higher than the experimental value of 3695 K. According to the calculations performed by Fikar et al., the melting points determined using the AT and DND potentials are approximately 5200 K and 3750 K, respectively [5]. The melting point using the extended FS potential by Dai et al. is 4500 K [26]. The melting points from the BOPs constructed by Juslin et al. [16] and by Li et al. [25] are 2750 K and 4500 K, respectively; The melting point using the MEAM potential is 4600 K [15]. Compared to most of the existing potentials, the present potential is able to reasonably reproduce the W melting point. Note that although the DND potential cannot correctly predict the thermal expansion, it can closely predict the melting point.

## 4. Defect properties

With the present potential, the vacancy migration energy that is not included in the fitting was calculated using the drag method. The resulting vacancy migration energy is 1.43 eV, which is very close to the value obtained using the AT potential (1.44 eV), but is lower than the experimental value of 1.7-2.02 eV. However, as indicated by Becquart and Domain [8], the comparison to experimental results is not straightforward, because the calculations were performed at 0 K, while the experimental data were obtained at high temperatures. Thus, the vacancy migration energy should be compared with the experimental values extrapolated to 0 K, i.e., 1.50 eV according to Satta et al. [27]. From this viewpoint, the vacancy migration energy is reasonable.

The properties of SIAs are of particular interest, because their production, accumulation, and migration will greatly affect the mechanical properties of materials. In this work, the diffusion behavior of SIAs was investigated using MD simulations. The observed migration of a $\langle 111 \rangle$ SIA, which is the most stable SIA configuration, is one dimensional in the $\langle 111 \rangle$ direction at temperatures below 550 K. For the convenience of analyses, here the atom that is located closest to the middle of two neighboring lattice sites is identified as a SIA. Fig. 3 illustrates the dynamics of $\langle 111 \rangle$ SIA diffusion in tungsten at 300 K. The atom S in Fig. 3a is an initial $\langle 111 \rangle$ SIA, and it displaces its neighboring atoms (A,B ,C and D) from their lattice sites. As the diffusion proceeds, one of the neighboring atoms (B or C) is displaced closest to the middle of two neighboring lattice sites and acts as a new SIA. This process is repeated, leading to the migration of a SIA. As shown in Fig. 3b, the atom G becomes a new SIA after 20 ps of system evolution. The atom S that is initially identified as a SIA occupies a lattice site (site b) that originally belonged to atom B. Except for the neighboring atoms of G that are displaced from their lattice sites, atoms between atom S and G replace in turn their nearest neighboring atoms in the $\langle 111 \rangle$ direction toward the position of atom G. Thus, we conclude that the initial SIA initiates a sequence of W atom displacements and replacements in $\langle 111 \rangle$ direction, and a new $\langle 111 \rangle$ SIA appears, accompanied by the annihilation of the old SIA.

For temperatures higher than 550 K, the migration proceeds in a three dimensional manner, due to the rotation of the $\langle 111 \rangle$ SIA configuration between three alternative crystallographically equivalent directions, for example, $[111]$, $[\bar{1}\bar{1}1]$, or $[\bar{1}11]$. As an example, Fig. 4 displays in detail this process at 550 K. The $[111]$ SIA configuration, which migrates one-dimensionally in the $[111]$ direction for most of the time, starts to rotate in the (110) surface when the system evolves for approximately 187 ps. After a system evolution of 1 ps, it transforms to the $[110]$ SIA configuration, which is an intermediate configuration for the event of rotation of the SIA configuration between the $[111]$ and $[11\bar{1}]$ directions. With further system evolution of approximately 1 ps, the $[11\bar{1}]$ SIA configuration forms, and it will continue to migrate one-dimensionally in the $[11\bar{1}]$ direction until the next rotation occurs. Similar phenomena have been observed in the results of MD simulations using the DND and AT potentials [5,19].

The diffusivity of SIAs in tungsten was studied by using the MD method on the basis of the Einstein relation given by:

$$D = \lim_{t \to \infty} \frac{\langle R(t)^2 \rangle}{2dt}, \qquad (6)$$

where $\langle R(t)^2 \rangle$ is the ensemble-averaged mean-square displacement of the SIA, $t$ is the diffusion time, and $d$ is the dimensionality of the space in which diffusion is occurring. The temperature of the system was varied between 100 K and 500 K. Depending on the temperature of the system, the simulation time was in the range of 2 ns to 7 ns. Because of efficiency considerations, the cubic simulation box was kept as small as possible with a side length of approximately 3.16 nm, containing 2000 tungsten atoms and one SIA. Note that, during the course of the simulation, as the SIA leaves the simulation box, one of its images will enter through the opposite face due to the periodic boundary conditions imposed. Consequently, the SIA always moves in the small simulation box, thus causing loss of information about the SIA diffusion However, as described in our previous work, the actual trajectory of the SIA in an infinite system can be achieved by correcting its position when a SIA migrates out of the simulation box [22].

Fig. 5 displays the SIA diffusion coefficient over the temperature range of 100-500 K. The diffusion data was fitted by an Arrhenius law equation $D = D_0 \exp(-E_A/k_B T)$, with $D_0 = 1.82 \times 10^{-7} \, \text{m}^2/\text{s}$ and $E_A = 0.022 \, \text{eV}$. In agreement with the other studies, this low migration energy indicates that SIAs in tungsten are very mobile. Recent molecular dynamics simulations by Derlet et al. using the DND potential indicate that the migration energy of a SIA in tungsten is 0.013 eV [19]. The experiments by Amano and Seidman [38] employing the filed-ion microscope technique demonstrated that SIAs underwent long-range migration and were mobile at temperatures as low as 28 K. The migration energy was found to be 0.079-0.085 eV. Dausinger and Schultz [39] concluded from their resistivity annealing experiments that SIA diffusion occurs between 24 K and 30 K, corresponding to a migration energy of 0.054 eV. More recently, Tamimoto et al. [40] proposed that SIA diffusion could already occur at temperatures below 1.5 K. Such a low migration temperature would correspond to a much lower migration energy than 0.054 eV. Thus, the migration energy of 0.022 eV determined in the present work is reasonable.

## 5. Conclusions

Due to the important influence of the W-W potential on the defect properties and the shortcomings of the currently available FS potential, in this paper, based on *ab initio* calculations, a modified Finnis-Sinclair-type interatomic potential for W-W interactions was developed. This potential is able to reproduce the correct SIA formation energies in tungsten, offering a significant improvement over the W-W potential by Ackland et al. Further, the modified potential was applied to study in detail the diffusion of a SIA in tungsten. An Arrhenius fit to the diffusion data determined the migration energy to be 0.022 eV, in reasonable agreement with the experimental data. The observed migration of a $\langle 111 \rangle$ SIA is one dimensional in the $\langle 111 \rangle$ direction at temperatures below 550 K. By investigating in detail the diffusion process, it is revealed that the initial SIA initiates a sequence of W atom displacements and replacements in the $\langle 111 \rangle$ direction, and a new $\langle 111 \rangle$ SIA appears that is accompanied by the annihilation of the old SIA. For temperatures higher than 550 K, the SIA migration begins to be three dimensional in character as a result of the rotation of the $\langle 111 \rangle$ SIA between three alternative crystallographically equivalent directions. Similar phenomena have been observed in the results of MD simulations using either the DND potential or the AT potential. It is shown that the modified potential developed in this paper can be used to investigate SIA properties and diffusion. Note that, similar to the AT and DND potentials, we do not include a short-range correction to this potential, which is necessary in studying high-energy collision dynamics. To simulate high-energy collision cascades, one can employ the approach presented by Fikar and Schaublin [5] to improve the reliability of the potential at short-range.

In addition, the thermal expansion, which is a fundamental physical property of materials, is calculated using this modified potential over a temperature range from 0 K to 3500 K, and the results are in reasonable agreement with the experimental data, thereby overcoming the shortcoming of the predicted negative thermal expansion using the W-W potential developed by Derlet et al. These results demonstrate that the present potential is valid over a fairly broad temperature range. We expect that the potential, which is constructed in such a way as to be able to retain a sufficiently high accuracy over a range of strongly distorted atomistic configurations and over a broad temperature range, will help understand the defect behavior and its effect on the microstructural and mechanical properties of tungsten.

**Acknowledgments**

The authors would like to thank Dr. Roman Gröger for stimulating comments. This work is partly supported by the National Natural Science Foundation of China (Grant Nos. 11175124 and 91126001) and the National Magnetic Confinement Fusion Program of China (2013GB109000).

**Figures**

**Fig. 1.** The present tungsten potential compared to the AT and DND potentials: (a) the electronic density function multiplied by $A^2$ and (b) the pairwise potential radial functions.

**Fig. 2.** Thermal expansion derived from the AT, DND and present potential in comparison with the experimental results.

**Fig. 3.** Typical diffusion process of SIA in tungsten at 300 K. The atoms labeled A-I are the tungsten atoms in the $\langle 111 \rangle$ string and are represented by the gray spheres. Labeled a-i are the corresponding lattice sites, which are represented by the black spheres. The atom S is the initial SIA. (a) t = 0 ps; (b) t = 20 ps.

**Fig. 4.** View of the SIA configuration consisting of the atoms A (gray sphere) and B (black sphere) in the $[010]$ direction. (a) t = 187 ps; (b) t = 187.8 ps; (c) t = 188 ps; (d) t = 188.2 ps; (e) t = 189 ps.

**Fig. 5.** Diffusion coefficient for a SIA in tungsten determined using molecular dynamics simulations and plotted as a function of the absolute temperature *T*. The solid line is the Arrhenius relation fitted to the corresponding data over the temperature range from 100 K to 500 K, with a migration energy of 0.022 eV and a prefactor of $1.82 \times 10^{-7} \, m^2/s$.

**Tables**

**Table 1** Comparison of the properties of bcc tungsten as obtained from experiment, *ab initio* calculations, the AT potential, the DND potential, and the modified FS potential derived in the present study. $E_c$: cohesive energy (eV/atom); $a$: lattice parameter (nm); $B$: bulk modulus (GPa); $C_{ij}$: elastic constants (GPa); $T_m$: melting point; $E_{vac}^f$ and $E_{vac}^m$ are the vacancy formation energy (eV) and the vacancy migration energy (eV), respectively; $E_{d111}^f$, $E_{d110}^f$, $E_{d100}^f$ are the formation energies (eV) of self-interstitial atoms with <111>, <110> and <100> configurations, respectively; $E_{tet}^f$ and $E_{oct}^f$ are the formation energies (eV) of a self-interstitial atom in tungsten for the tetrahedral and octahedral sites, respectively.

|  | Present study | *Ab initio* | Experiment | DND[a] | AT[b] |
|---|---|---|---|---|---|
| $E_c$ | -8.9 | -9.97[c], -8.49[d] | -8.9[e] | -8.9 | -8.9 |
| $a$ | 0.3165 | 3.14[c], 3.172[d] | 0.3165[f] | 0.3165 | 0.3165 |
| $B$ | 309 | 320[g] | 308-314[h] | 310 | 310 |
| $C_{11}$ | 520 | 552[g] | 501-521[h] | 525 | 522 |
| $C_{12}$ | 204 | 204[g] | 199-207[h] | 203 | 204 |
| $C_{44}$ | 161 | 149[g] | 151-160[h] | 159 | 161 |
| $E_{vac}^{f}$ | 3.58 | 3.56[i], 3.11[j] | 3.1-4.0[k,l] | 3.56 | 3.63 |
| $E_{vac}^{m}$ | 1.43 | 1.78[i], 1.66[j] | 1.7-2.0[m,n] | 2.07[p] | 1.44[p] |
|  |  |  | 1.5 at 0 K[j,o] |  |  |
| $E_{d111}^{f}$ | 9.58 | 9.55[i], 9.82[j] |  | 9.55 | 8.92 |
| $E_{d110}^{f}$ | 9.86 | 9.84[i], 10.10[j] |  | 9.84 | 9.64 |
| $E_{d100}^{f}$ | 11.53 | 11.49[i], 11.74[j] |  | 11.51 | 9.82 |
| $E_{tet}^{f}$ | 10.93 | 11.05[i], 11.64[j] |  | 11.00 | 10.00[p] |
| $E_{oct}^{f}$ | 11.72 | 11.68[i], 11.99[j] |  | 11.71 | 10.02[p] |
| $T_m$ | 4100±50 |  | 3695 | 3750[q] | 5150-5250[q] |

[a]Reference 19.

[b]Reference 18.

[c]Reference 28.

[d]Reference 29.

[e]Reference 30.

[f]Reference 31.

[g]Reference 32.

[h]Reference 33.

[i]Reference 10.

[j]Reference 8.

[k]Reference 34.

[l]Reference 35.

[m]Reference 36.

[n]Reference 37.

[o]Reference 27.

[p]Present work.

[q]Reference 5.

**Table 2** Parameters for the present W-W potential. The units are combinations of eV and Å to have Eqs. (1) to (5) be expressed in units of eV.

| $A_n$ | $B_n$ | $C_n$ | $D_n$ |
|---|---|---|---|
| 836.733119131673 | -1092.73723146209 | 570.244952195581 | -149.176023687989 |
| 185.476611562764 | -65.1531905705160 | -144.143194717123 | 128.381692972279 |
| 4483.32942908444 | 798.771513881627 | -3119.64856577141 | -2188.34199479366 |
| 497.885673416326 | -663.034572443750 | 332.677312093750 | -74.4188648500000 |
| 1033.67091988025 | -1124.85381286782 | 408.726035584526 | -49.5696850121021 |
| 10782.4399433266 | -13005.0298918015 | 5234.58604338040 | -703.009068965912 |

Potential parameters.

| $E_n$ | $F_n$ | $G_n$ | $H_n$ | $r_n$ |
|---|---|---|---|---|
| 19.7590956539449 | -1.10929154438237 | 8.583068847656250E-3 | 0.0 | 4.25270034968853 |
| -42.2614091514405 | 6.18967940283576 | -0.329017639160156 | 0.0 | 3.18798422753811 |
| 3852.65707605729 | -1773.50698078752 | 352.122068405151 | -26.2997150421143 | 3.08288478612900 |
| 6.25419990000000 | 0.0 | 0.0 | 0.0 | 2.83075976371765 |
| 0.0 | 0.0 | 0.0 | 0.0 | 2.72917079940438 |
| 0.0 | 0.0 | 0.0 | 0.0 | 2.46177387237549 |

Potential parameters.

| $A'_n$ | $B'_n$ | $C'_n$ | $D'_n$ |
|---|---|---|---|
| 0.176113968245774 | -4.42727651433102 | 6.08596367085100 | -3.31627185619130 |
| -9.01097257391517 | 17.7652430017672 | -14.0448182382562 | 5.85262981750376 |
| 2941.47663526372 | -4976.29995214059 | 3363.92566298462 | -1135.68505574601 |
| -506.899715533763 | 1117.19138346861 | -1025.00865314071 | 501.406917325721 |
| 170.584208417048 | -191.839109698533 | 72.0025483634399 | -9.01365494002994 |
| 98.5798682207654 | -105.456860917279 | 37.4588301923518 | -4.40904970411985 |

Potential parameters.

| $E'_n$ | $F'_n$ | $G'_n$ | $H'_n$ | $r'_n$ |
|---|---|---|---|---|
| 0.889234910474537 | -0.117663809722679 | 6.163730456134172E-3 | 0.0 | 4.44999316573143 |
| -1.36849120887594 | 0.170647289763123 | -8.865704319423323E-3 | 0.0 | 3.26985955238342 |
| 191.481369497020 | -12.8985156307958 | 0.0 | 0.0 | 3.14614844381809 |
| -137.946827388616 | 20.2386766904903 | -1.23706225652103 | 0.0 | 2.79194164276123 |
| 0.0 | 0.0 | 0.0 | 0.0 | 2.66148281335831 |
| 0.0 | 0.0 | 0.0 | 0.0 | 2.50066447257996 |

Potential parameters.

| $N$ | $N'$ | $r_{N+1}$ | $r'_{N+1}$ | $A$ |
|---|---|---|---|---|
| 6 | 6 | 0.0 | 0.0 | 10.3632760643959 |

Figure 1

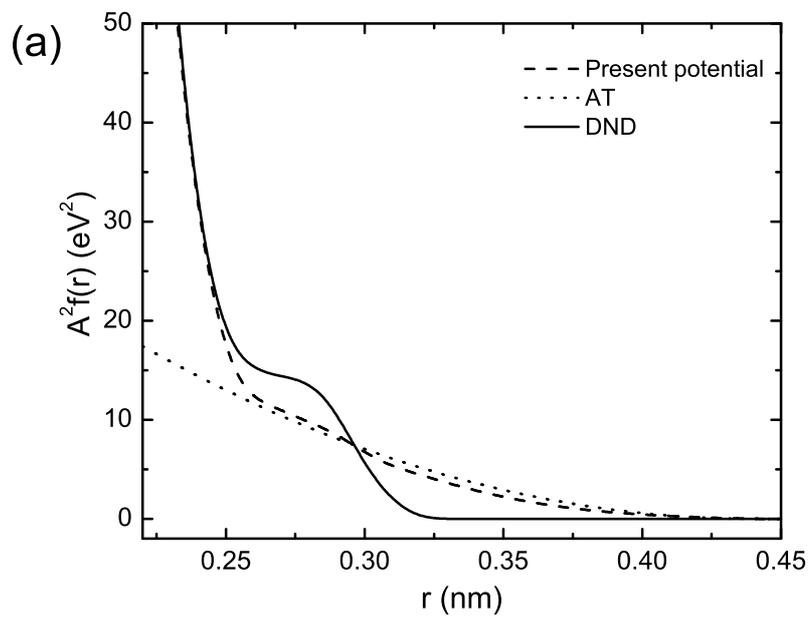

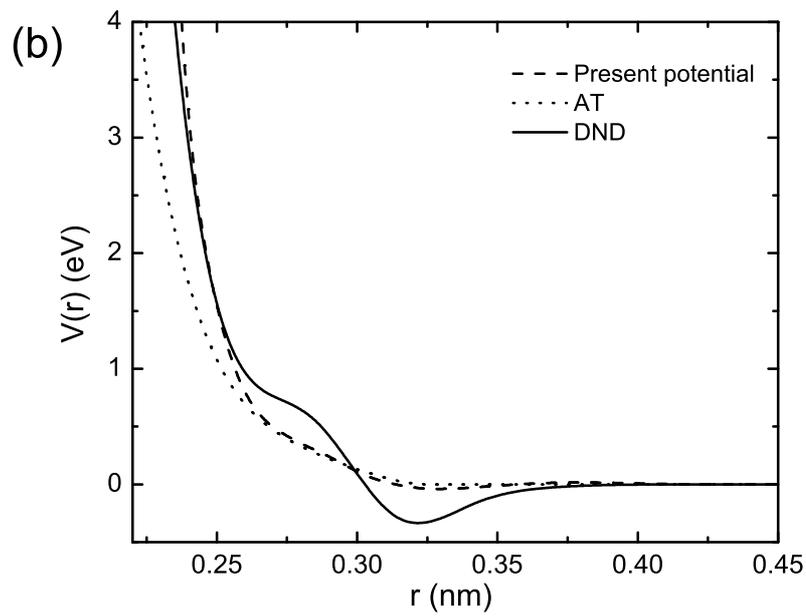

Figure 2

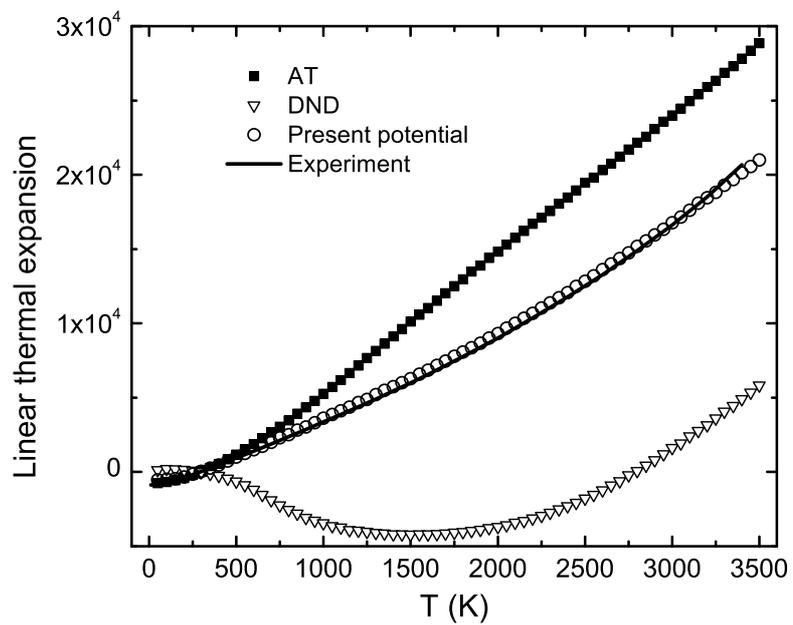

Figure 3

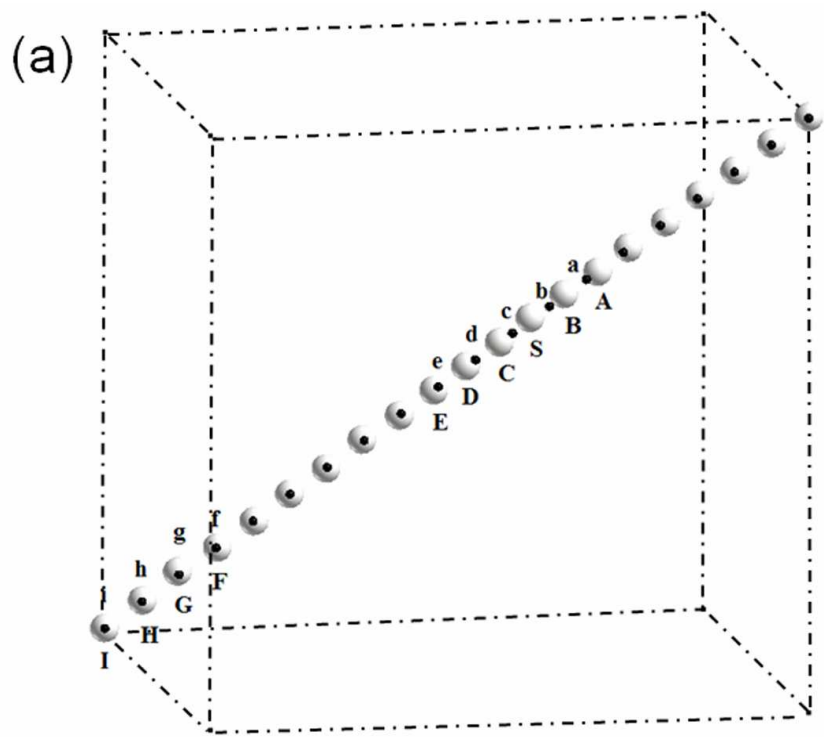

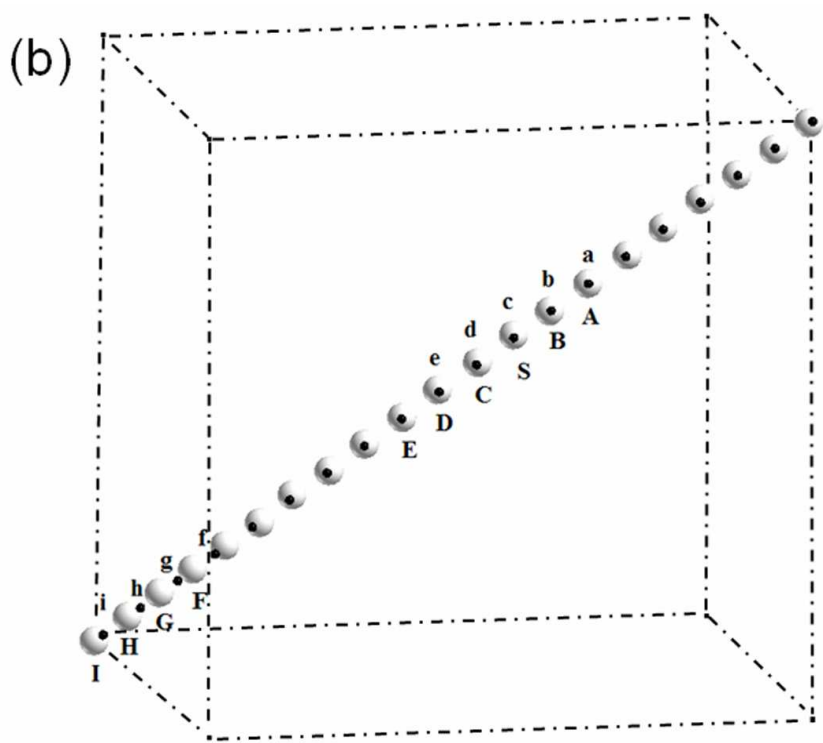

Figure 4

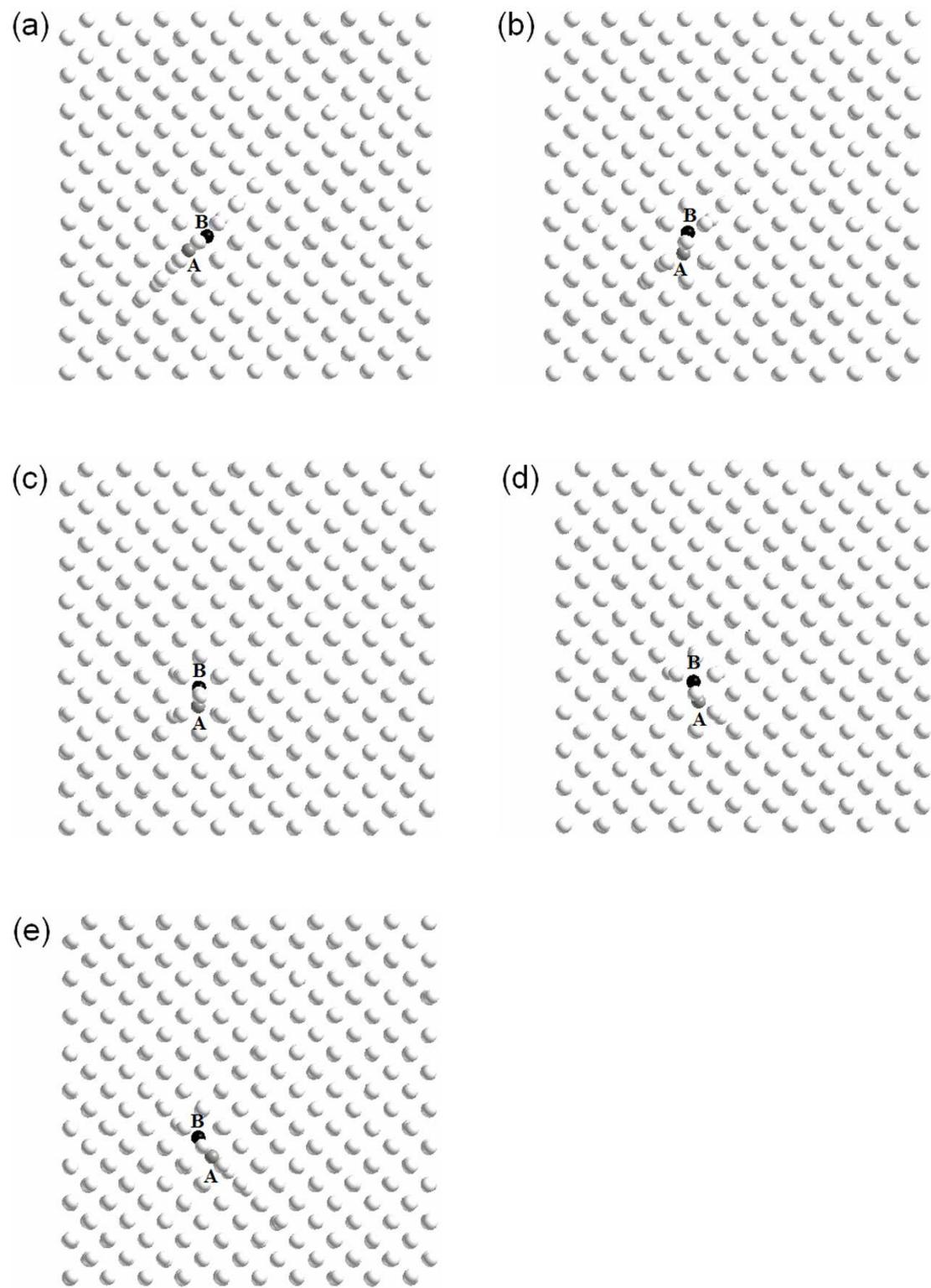

Figure 5

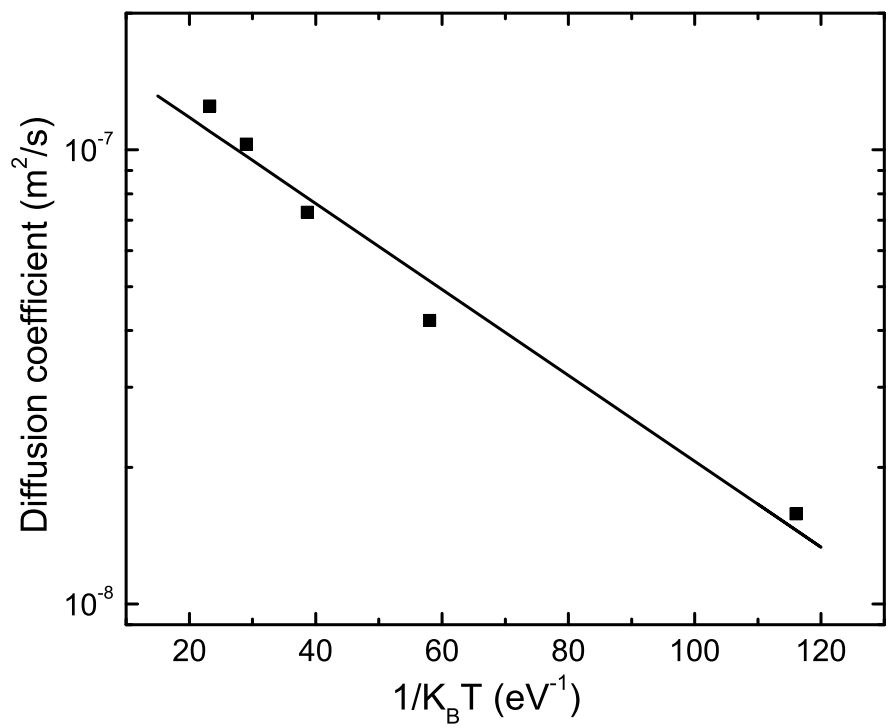